\documentclass[twocolumn,aps,superscriptaddress,prb]{revtex4-1}
\setcitestyle{numbers,square}
\usepackage{amsmath,amssymb}
\usepackage{tabularx}
\usepackage{graphicx}% Include figure files \usepackage{dcolumn}% Align table columns on decimal point
\usepackage{bmpsize}
\usepackage{bm}% bold math
\usepackage{color}
\usepackage{amssymb}
\usepackage[version=3]{mhchem}
\usepackage{newtxmath}
\DeclareMathAlphabet{\mathpzc}{OT1}{pzc}{m}{it}
\newcommand{\nn}{\nonumber}

\def\i{\mathrm{i}}
\def\e{\mathrm{e}}

\begin{document}
%\preprint{APS/123-QED}
%%%%%%%%%%%%%%%%%%%%%%%%%%%%%%%%%%%%%%%%%%%%%%%%%%%%%%%%%%%%%%%%%%%%%%%%%%%%%%%%%%%%%%%%%%%%%%%%%%%%%%%%%%%%%%%%%%%%%%%
\title{
Moir\'e disorder effect in twisted bilayer graphene
}
\author{Naoto Nakatsuji}
\affiliation{Department of Physics, Osaka University, Toyonaka, Osaka 560-0043, Japan}
\author{Mikito Koshino}
\affiliation{Department of Physics, Osaka University, Toyonaka, Osaka 560-0043, Japan}
\date{\today}
%%%%%%%%%%%%%%%%%%%%%%%%%%%%%%%%%%%%%%%%%%%%%%%%%%%%%%%%%%%%%%%%%%%%%%%%%%%%%%%%%%%%%%%%%%%%%%%%%%%%%%%%%%%%%%%%%%%%%%%

%%%%%%%%%%%%%%%%%%%%%%%%%%%%%%%%%%%%%%%%%%%%%%%%%%%%%%%%%%%%%%%%%%%%%%%%%%%%%%%%%%%%%%%%%%%%%%%%%%%%%%%%%%%%%%%%%%%%%%%
\begin{abstract}
We theoretically study the electronic structure of magic-angle twisted bilayer graphene with disordered moir\'e patterns.
By using an extended continuum model incorporating non-uniform lattice distortion, we find that the local density of states of the flat band is hardly broadened, but splits into upper and lower subbands in most places. 
The spatial dependence of the splitting energy is 
almost exclusively determined by the local value of the effective vector potential induced by heterostrain,
whereas the variation of local twist angle and local moir\'e period give relatively minor effects on the electronic structure.
We explain the exclusive dependence on the local vector potential by a pseudo Landau level picture for the magic-angle flat band, and we obtain an analytic expression of the splitting energy as a function of the strain amplitude.
%We theoretically study the electronic structure of magic-angle twisted bilayer graphene with general disordered moir\'e patterns which some experiments observed recently.
%we calculate the energy spectrum by using an extended continuum model incorporating lattice distortion.
%We find that the local density of states (LDOS) of the flat band is hardly broadened but splits place by place. Remarkably, the spatial variation of the splitting energy is totally uncorrelated with the local twist angle, but it is almost exclusively determined by the local value of the effective vector potential caused by heterostrain (relative strains between layers). 
%We explain the exclusive dependence on the strain-induced vector potential by using a pseudo landau level picture for the magic-angle flat band, and obtain an analytic expression for the splitting energy as a function of the strain amplitude.
\end{abstract}
%%%%%%%%%%%%%%%%%%%%%%%%%%%%%%%%%%%%%%%%%%%%%%%%%%%%%%%%%%%%%%%%%%%%%%%%%%%%%%%%%%%%%%%%%%%%%%%%%%%%%%%%%%%%%%%%%%%%%%%

\maketitle
%%%%%%%%%%%%%%%%%%%%%%%%%%%%%%%%%%%%%%%%%%%%%%%%%%%%%%%%%%%%%%%%%%%%%%%%%%%%%%%%%%%%%%%%%%%%%%%%%%%%%%%%%%%%%%%%%%%%%%%
\section{introduction}
\label{sec:intro}

Twisted bilayer graphene (TBG) exhibits various exotic quantum phenomena with a wide variety of correlated phases
\cite{Cao2018_80,cao2018_43,doi:10.1126/science.aav1910,Kerelsky2019,xie2019spectroscopic,jiang2019charge,Choi2019,doi:10.1126/science.aaw3780,polshyn2019large,lu2019superconductors,PhysRevLett.124.076801,doi:10.1126/science.aay5533,chen2020tunable,saito2020independent,zondiner2020cascade,wong2020cascade,stepanov2020untying,arora2020superconductivity,PhysRevLett.127.197701}.
%such as correlated insulating phases\cite{Cao2018_80}, superconductivity\cite{cao2018_43,doi:10.1126/science.aav1910} and anomalous Hall effect\cite{doi:10.1126/science.aaw3780,doi:10.1126/science.aay5533,PhysRevLett.127.197701}. 
These quantum states originate from moir\'e-induced flat bands, which emerge when two graphene layers stacked with a magic-angle ($\sim 1^\circ$)\cite{PhysRevB.82.121407,trambly2010localization,doi:10.1073/pnas.1108174108,PhysRevB.86.125413}.
The flat band is usually described by a theoretical model assuming a regular moir\'e superlattice with a perfect periodicity \cite{trambly2010localization,doi:10.1073/pnas.1108174108,PhysRevB.82.121407,PhysRevB.86.125413,PhysRevB.86.155449,PhysRevX.8.031087,kang2018symmetry,PhysRevX.8.031089,doi:10.1073/pnas.1810947115,PhysRevX.10.031034,PhysRevLett.124.097601,PhysRevB.102.035136,PhysRevB.103.035427,PhysRevB.87.205404,koshino2015interlayer,carr2020electronic}.
However, 
the moir\'e interference pattern is highly sensitive to a slight distortion of underlying structure.
In TBG, an atomic displacement of graphene's lattice is magnified in the moir\'e superlattice by factor of the inverse twist angle \cite{cosma2014moire},
leading to unavoidable disorder in the moir\'e superlattice.
Indeed, the moir\'e patterns in actual TBG samples are not perfectly regular, but exhibit non-uniform structures including local distortion and variance of the twist angle 
\cite{xie2019spectroscopic,jiang2019charge,Choi2019,Kerelsky2019,li2010observation,PhysRevLett.106.126802,PhysRevLett.109.196802,PhysRevB.92.155409,PhysRevB.98.235402,yoo2019atomic,doi.10.1038/s41467-019-14207-w,Nature.581.7806,McGilly2020,gadelha2021localization,kazmierczak2021strain,tilak2021flat,PhysRevLett.127.126405,huang2021moir,schapers2021raman}.

%However, the moir\'e interference pattern is highly sensitive to a slight distortion of underlying structure, such that an atomic displacement of graphene's lattice is magnified by the factor of $\theta^{-1}$ in the moir\'e pattern, where $\theta$ is the twist angle.\cite{cosma2014moire}
%Indeed, the moir\'e pattern in actual TBG samples is not perfectly regular, but exhibits a non-uniform structure including a local variance of the twist angle \cite{Nature.581.7806,PhysRevB.98.235402,Kerelsky2019,McGilly2020,Choi2019,doi.10.1038/s41467-019-14207-w}.

It is expected that such a disorder in the moir\'e pattern would strongly affect the flat band and its electronic properties in the one-body level.
Generally, non-uniform moir\'{e} systems are hard to treat theoretically, because one needs to consider a number of moir\'e periods each of which contains huge number of atoms.
In previous works, the effect of the twist angle disorder in TBG was investigated using various theoretical approaches, 
such as a real-space domain model composed of regions with different twist angles \cite{PhysRevResearch.2.023325},
transmission calculations through one-dimensional variation of twist angle \cite{PhysRevResearch.2.033458,joy2020transparent,PhysRevB.104.075144}, and a Landau-Ginzburg theory to study the interplay between electron-electron interactions and disorder \cite{PhysRevB.103.125138}.

%On the other hand, the effect of uniformly-distorted TBG was investigated in some previous works, and it was shown that heterostrain
%(relative strains between layers) leads to an energy separation of the flat band.\cite{PhysRevB.100.035448, PhysRevLett.120.156405,PhysRevB.98.235402}

\begin{figure}[t]
  \begin{center}
    \leavevmode\includegraphics[width=1. \hsize]{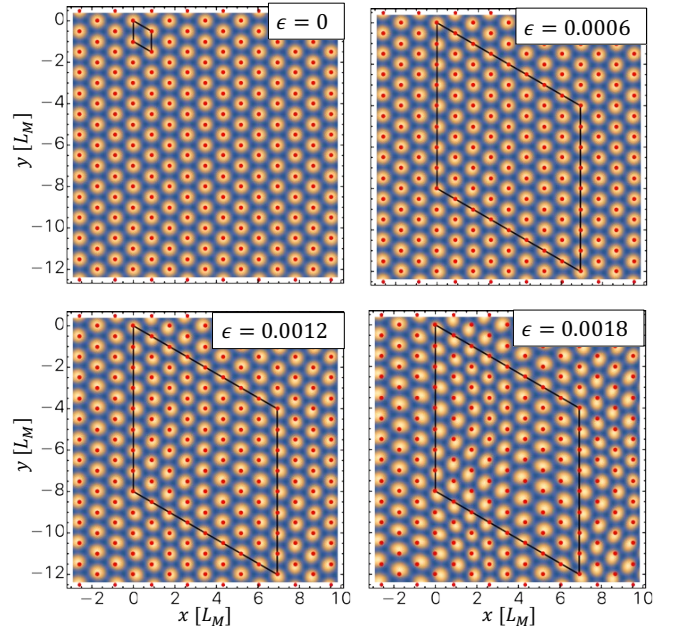}
    \caption{
    Moir\'e patterns of magic-angle TBG $(\theta = 1.05^\circ)$ with random non-uniform distortion of $\epsilon = 0$, $0.0006$, $0.0012$ and $0.0018$,
    where the characteristic wave length is $\lambda = 7L_{M}$,
    and the super-cell size (big parallelogram) is $n_{\rm SM} = 8$.
    The bright region represents local AA stack and the dark region represents AB/BA stack. The red dots are the AA spots of the non-distorted TBG for reference.
      }
    \label{Fig3_Lattice_distortion}
  \end{center}
  \end{figure}
  
In this paper, we study the electronic structure of magic-angle TBG in the presence of  non-uniform moir\'e patterns as shown in Fig.~\ref{Fig3_Lattice_distortion},
generated from random lattice distortion of graphene layers.
%to simulate realistic situations.
The model automatically contains possible moir\'e disorder components, including various types of local strains and local rotations.
%We assume a $n_{\rm SM}\times n_{\rm SM}$ super-moir\'e structure 
We calculate the energy spectrum by using an extended continuum model incorporating non-uniform lattice distortion \cite{PhysRevB.101.195425}.
We find that the local density of states (LDOS) of the flat band is hardly broadened but splits place by place. Remarkably, the spatial variation of the splitting energy is totally uncorrelated with local twist angle or local periodicity, but it is almost exclusively determined by the local value of the effective vector potential caused by heterostrain, or relative strains between layers. 
We explain the exclusive dependence on the strain-induced vector potential by using a pseudo Landau level picture for the magic-angle flat band \cite{PhysRevB.99.155415}, and obtain an analytic expression for the splitting energy as a function of the strain amplitude.
The strain-induced flat band splitting is an analog of that in uniformly-distorted TBGs
\cite{PhysRevLett.120.156405,PhysRevB.100.035448,he2020giant,xie2019spectroscopic,PhysRevB.104.045403,PhysRevResearch.4.013209,PhysRevB.102.201107,PhysRevLett.127.027601},
and the strong coincidence between the splitting energy and the local strain tensor in non-uniform TBGs reflects a highly-localized feature of the flat band wave function.

This paper is organized as follows.
Before we consider non-uniform moir\'e disorder,
we present in Sec.~\ref{sec_um} a detailed study on a TBG with uniform distortion.
We investigate the effects of different types of strain components independently, and show that the flat band splitting is mainly caused by shear and anisotropic-normal heterostrain. We derive an approximate expression for the splitting energy by using the pseudo Landau level analysis.
%In Sec.~\ref{sec_um}, we consider a magic-angle TBG with uniform distortion and investigate the effects of different types of strain components independently. We demonstrate that the flat band splitting is mainly caused by shear and anisotropic-normal heterostrain. We derive an approximate expression for the splitting by using pseudo Landau level analysis.
In Sec.~\ref{sec:n-um}, we calculate the LDOS
of magic-angle TBG with non-uniform moir\'e patterns,
and demonstrate a strong relationship between the LDOS split and the strain-induced vector potential.
A brief conclusion is given in Sec.~\ref{sec_con}.

\section{TBG with a uniform distortion}
\label{sec_um}
 \subsection{Atomic structure}
	%%% \subsubsection{Atomic structure}
	We first consider a TBG with a uniform lattice distortion
	and investigate its effect on the flat band.
	We define the lattice vectors of monolayer graphene as $\bm{a}_{1}=a(1, 0)$ and $\bm{a}_{2} = a(1/2, \sqrt{3}/2)$ where
	$a = 0.246$ nm is the lattice constant,
	and define $\bm{b}_{j}$ as the corresponding reciprocal lattice vectors to satisfy $\bm{a}_i\cdot\bm{b}_j=\delta_{ij}$.
	In a perfect TBG  without distortion,
	the lattice vectors of layer $l (= 1,2)$ are given by
	$\bm{a}_{j}^{(l)} = R(\mp\theta/2)\bm{a}_{j}$
	where $\mp$ is for $l = 1$ and $2$, respectively,
	$R$ is a two-dimensional rotation matrix, and $\theta$ is the twist angle.
	
We introduce a uniform distortion to layer $l$, whch is expressed by a matrix,
\begin{equation}
    \mathcal{E}^{(l)}
    = \begin{pmatrix}
		\epsilon^{(l)}_{xx} & - \Omega^{(l)}+\epsilon^{(l)}_{xy}  \\
		\Omega^{(l)}+\epsilon^{(l)}_{xy} & \epsilon^{(l)}_{yy}
				\end{pmatrix}.
\end{equation}
The $\mathcal{E}^{(l)}$ represents a deformation such that a carbon atom at a position $\bm{r}$ in a non-distorted system is shifted to $\bm{r} + \mathcal{E}^{(l)}\bm{r}$.
Here $\epsilon_{xx}^{(l)}$ and $\epsilon_{yy}^{(l)}$ 
represent normal strains in $x$ and $y$ directions, respectively,
$\epsilon_{xy}^{(l)}$ is a shear strain,
and $\Omega^{(l)}$ is a rotation from the original twist angle 
For later arguments, we also define the isotropic/anisotropic components of the normal strain by
\begin{equation}\label{eq_epsilon_pm}
    \epsilon_{\pm}^{(l)} = 
    \frac{1}{2}(\epsilon_{xx}^{(l)} \pm \epsilon_{yy}^{(l)}),
\end{equation}
and the interlayer difference of each strain/rotation component as
    \begin{align}\label{eq_Delta_eps}
    \epsilon_{\pm} &= \epsilon_{\pm}^{(1)}-\epsilon_{\pm}^{(2)},
    \notag\\
    \epsilon_{xy} &= \epsilon_{xy}^{(1)}-\epsilon_{xy}^{(2)},
    \notag\\
    \Omega &= \Omega^{(1)}-\Omega^{(2)}.
    \end{align}

In the presence of distortion, the lattice vectors  
change to $\bm{a}_{j}^{(l)}=(1+\mathcal{E}^{(l)})R(\mp\theta/2)\bm{a}_{j}$.
In the following, we assume the original twist angle and the distortion is sufficiently small ($\theta, \Omega^{(l)}, \epsilon^{(l)}_{\mu\nu} \ll 1$), so that
	\begin{align}
	\bm{a}_{j}^{(l)} \approx \bigl[ R\left(\mp\theta/2\right)+\mathcal{E}^{(l)} \bigr]
	\bm{a}_{j}.
	\end{align}
Similarly, the reciprocal lattice vectors are written as 
	\begin{align}
	\bm{b}_{j}^{(l)} 
	%&= \Bigl[\left( R\left(\mp\theta/2\right)+\mathcal{E}^{(l)} \right)^{-1}\Bigr]^{\rm T} \bm{b}_{j}
	%\notag\\
	& \approx \bigl[ R\left(\mp\theta/2\right)- \mathcal{E}^{(l)\rm T} \bigr] \bm{b}_{j},
	\end{align}
where $T$ is the matrix transpose.

\begin{figure}
  \begin{center}
    \leavevmode\includegraphics[width=1. \hsize]{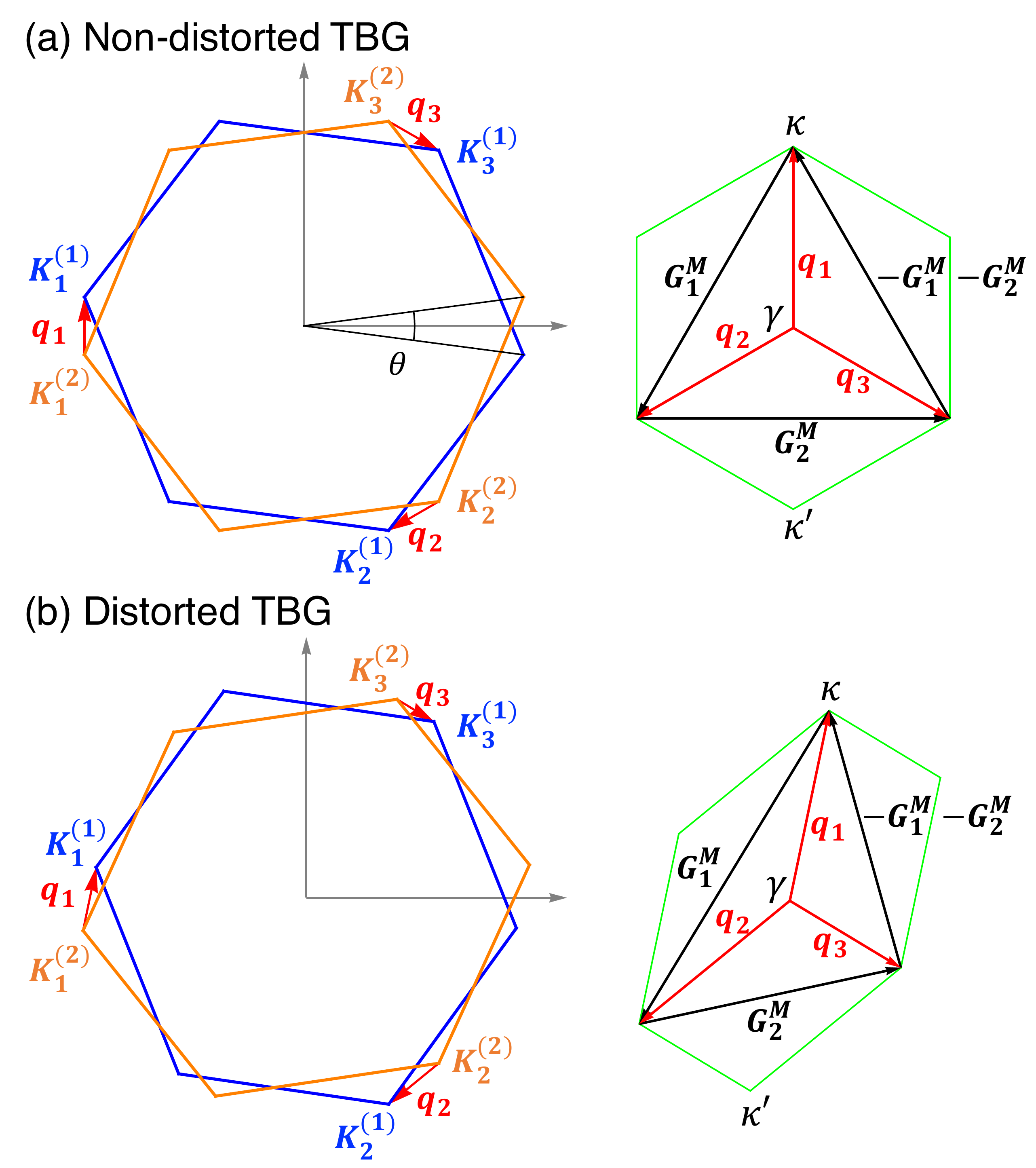}
    \caption{
    Brillouin zones of (a) a non-distorted TBG and (b) a distorted TBG.
    Blue and orange hexagons on the left represent the first Brillouin zone of graphene layer 1 and 2 (twisted by $\mp \theta/2$), respectively, and 
    red arrows are the displacement vectors from the layer 2's $K_+$ point to layer 1's.
    A green hexagon on the right side is the moir\'{e} Brillouin zone.
%
%    (a) The Brillouin zone(BZ) of each layers and $\bm{q}$ vector without moir\'e distortion.
%   	Blue line is layer1's BZ, and Orange is layer2's respectively.
%	these are twisted $\mp \theta/2$.
%	Red arrows show that the vector from layer2's K point to layer1's (we draw only ($\xi = +$)-valley).
%    (b)The moir\'e BZ without distortion.
%    	Red arrows are vectors which give the difference between K point of each layers as (a).
%	Black arrows ($\bm{G}_{1}^{M}, \bm{G}_{2}^{M}$) are moir\'e reciprocal lattice vectors.
%    (c) The mor\'e BZ with distortion.
%    	As the case without distortion(b), we can define the $\bm{q}_{j}$ and $\bm{G}_{j}^{M}$ vectors.
    }
    \label{Fig1_BZ}
  \end{center}
  \end{figure}

In an intrinsic monolayer graphene,
six corner points of the Brillouin zone (BZ) are given by $\xi \bm{K}_j\, (j=1,2,3)$, where $\xi = \pm 1$ label the valley degree of freedom, and 
\begin{align}\label{eq_K_graphene}
	\bm{K}_{j} &= R\left(\phi_{j}\right)\frac{4\pi}{3a}(-1,0),\quad
	\phi_j = \frac{2\pi}{3}(j-1),
\end {align}
are equivalent points in the BZ.
Corresponding vectors for the disroted TBG are written as 
	\begin{align}
	\bm{K}_{j}^{(l)} 
	& \approx \bigl[ R\left(\mp\theta/2\right)- \mathcal{E}^{(l)\rm T} \bigr] \bm{K}_{j}.
	\end{align}
Figure \ref{Fig1_BZ} illustrates the schematics of BZ
for (a) a non-distorted TBG and (b) a distorted TBG.
In each panel, blue and orange hexagons on the left represent the first BZ of graphene layer $l=1$ and 2, respectively,
where the corner points are given by 
$\xi \bm{K}_{j}^{(l)}$.
We define interlayer shift of the corner points by
	\begin{equation}\label{eq_q}
	    \bm{q}_j = \bm{K}_{j}^{(1)} - \bm{K}_{j}^{(2)}
	    \quad (j=1,2,3),
	\end{equation}
	as shown in Fig.~\ref{Fig1_BZ}.
The $\bm{q}_j$'s can be expressed only by the interlayer rotation and strain components as
\begin{align}\label{eq_q2}
    \bm{q}_{j} &= 
    \frac{4\pi}{3a}\left[
    R\left(\phi_{j}\right)
    \begin{pmatrix}
    \epsilon_{+} \\ \theta-\Omega
    \end{pmatrix}
    + R\left(-\phi_{j}\right)
    \begin{pmatrix}
    \epsilon_{-} \\ \epsilon_{xy}
    \end{pmatrix}
    \right].
\end{align}

%where $\phi_j = (2\pi/3)(j-1)$. 

  %  Figure \ref{Fig1_BZ} illustrates the schematics of the Brillouin zone (BZ)
  %  for (a) a non-distorted TBG and (b) a distorted TBG.
  %  In each panel, blue and orange hexagons on the left represent the first BZ of graphene layer 1 and 2, respectively.
  %  The Dirac points of layer $l$ are located at corners of the hexagon, or 
%	\begin{align}
%	\bm{K}_{\xi}^{(l)} &= - \xi (2 \bm{b}_{1}^{(l)} + \bm{b}_{2}^{(l)})/3,
%	\end {align}
%	where $\xi = \pm 1$ labels the valley degree of freedom.
%	Three corner points for valley $\xi$ are given by
%	\begin{align}\label{eq_K_123}
%	    \bm{K}_{\xi,1}^{(l)} &= \bm{K}_{\xi}^{(l)}, 
%	    \notag\\
%	    \bm{K}_{\xi,2}^{(l)} &= \bm{K}_{\xi}^{(l)}+\xi \bm{b}^{(l)}_1, \notag\\ 
%	    \bm{K}_{\xi,3}^{(l)} &= \bm{K}_{\xi}^{(l)}+\xi (\bm{b}^{(l)}_1+\bm{b}^{(l)}_2),
%	\end{align}
%	which are equivalent points in the BZ.
%	We define interlayer shift of the corner points for $\xi=+$ by
%	\begin{equation}\label{eq_q}
%	    \bm{q}_j = \bm{K}_{+,j}^{(1)} - \bm{K}_{+,j}^{(2)}
%	    \quad (j=1,2,3),
%	\end{equation}
%	as shown in Fig.~\ref{Fig1_BZ}.

The reciprocal lattice vectors of the moir\'e pattern are given by $\bm{G}^{M}_{j} = \bm{b}^{(1)}_{j} - \bm{b}^{(2)}_{j}$,
which are also written as
$\bm{G}^{\rm M}_{1} = \bm{q}_2 - \bm{q}_1$, $\bm{G}^{\rm M}_{2} = \bm{q}_3 - \bm{q}_2$. In Fig.~\ref{Fig1_BZ}, a green hexagon on the right side represents the moir\'{e} Brillouin zone defined by $\bm{G}^{\rm M}_{j}$'s

 \subsection{Continuum model and Band calculation}
 %for TBG \\with uniform strain
    We use the continuum model 
    	\cite{PhysRevB.87.205404,koshino2015interlayer,doi:10.1073/pnas.1108174108,PhysRevB.86.155449,doi:10.1073/pnas.1810947115,PhysRevX.8.031087,kang2018symmetry,PhysRevX.8.031089,PhysRevB.100.035448,PhysRevX.10.031034,PhysRevLett.124.097601,PhysRevB.102.035136,PhysRevB.103.035427,PhysRevB.104.045403,PhysRevResearch.4.013209,he2020giant,PhysRevB.101.195425, PhysRevB.99.155415, PhysRevLett.120.156405,PhysRevB.100.035448,he2020giant,PhysRevB.104.045403,PhysRevResearch.4.013209,PhysRevB.102.201107,PhysRevLett.127.027601,carr2020electronic,guinea2019continuum,carr2019exact} 
    to describe a strained TBG.
	%We use the continuum model\cite{PhysRevB.100.035448} of TBG to calculate band structure included lattice distortion.
  %	Each valley of monolayer graphene are far away in the k-space when rotation angle and distortion elements are small enough, so we ignore the mixing between these valley.
%	thus, we split Hamiltonian of TBG into each valleys completely.
	The effective Hamiltonian for valley $\xi$ 
	is written as

	\begin{align} \label{eq:TBGHamiltonian}
	{\cal H}^{(\xi)}(\bm{k}) = \left(
		\begin{array} {cc}
		H_{1}(\bm{k}) & U^{\dagger} \\
		U & H_{2}(\bm{k})
		\end{array}
		\right),
	\end{align}
	where $H_l(\bm{k})$ is the $2\times 2$ Hamiltonian of distorted monolayer graphene,
	and $U$ is the interlayer coupling matrix.
	The Hamiltonian[Eq.~\eqref{eq:TBGHamiltonian}] works on
	the four-component wave function  $(\psi_{A}^{(1)}, \psi_{B}^{(1)}, \psi_{A}^{(2)}, \psi_{B}^{(2)})$, 
	where $\psi_{X}^{(l)}$ represents the envelope function of sublattice $X(=A,B)$ on layer $l(=1,2)$. 
	%I use the basis $(\psi_{A}^{(1)}, \psi_{B}^{(1)}, \psi_{A}^{(2)}, \psi_{B}^{(2)})$.
	
	The $H_l(\bm{k})$ is given by
	\begin{flalign} \label{eq:monolayerGrapheneHamiltonian}
	H_{l}(\bm{k})=-\hbar v\left[\left(R\left(\mp\theta\right)+\mathcal{E}^{(l)} \right)^{-1} \left(\bm{k}+\frac{e}{\hbar}\bm{A}^{(l)}\right)\right]\cdot \bm{\sigma},
	\end{flalign}
	where $\mp$ is for $l = 1$ and $2$, respectively, $v$ is the graphene's band velocity, $\bm{\sigma} = \left(\xi \sigma_{x}, \sigma_{y}\right)$ and $\sigma_{x}$, $\sigma_{y}$ are the Pauli matrices in the sublattice space $\left(A, B\right)$.
	%We take $\hbar v/a = 2.1354$ eV. 
	We take $\hbar v/a = 2.14$ eV \cite{PhysRevX.8.031087}. 
	The $\bm{A}^{(l)}$ is the strain-induced vector potential that is given by \cite{PhysRevB.65.235412,PhysRevLett.103.046801,Guinea2010}
	\begin{eqnarray}\label{eq_A1}
	\bm{A}^{(l)} &=& 
		\xi \frac{3}{2}\frac{\beta\gamma_{0}}{ev}
	\begin{pmatrix}
	\epsilon_{-}^{(l)}\\
	-\epsilon_{xy}^{(l)}\
	\end{pmatrix},
%	\xi \frac{3}{4}\frac{\beta\gamma_{0}}{ev}
%	\begin{pmatrix}
%	\epsilon_{xx}^{(l)}-\epsilon_{yy}^{(l)}\\
%	-2\epsilon_{xy}^{(l)}\
%	\end{pmatrix},
	\end{eqnarray} 
	where $\gamma_{0}=2.7$ eV is the nearest neighbor transfer energy of intrinsic graphene and $\beta \approx 3.14$.
%	We define the isotropic/anisotropic components of the normal strain as
%	\begin{equation}
%	    \epsilon_{\pm}^{(l)} \equiv 
%	    \frac{1}{2}(\epsilon_{xx}^{(l)} \pm \epsilon_{yy}^{(l)}).
%	\end{equation}
    Note that the strain-induced vector potential, Eq.~\eqref{eq_A1}, depends only on $\epsilon_{-}^{(l)}$ and $\epsilon_{xy}^{(l)}$,
    while not on $\epsilon_{+}^{(l)}$ or $\Omega^{(l)}$.

	The interlayer coupling matrix $U$ is given by
	\begin{align} \label{eq:Moire_interaction}
	& U= \sum_{j=1}^{3} U_{j} \e^{\i \xi\bm{q}_{j} \cdot \bm{r}}, \nn
%	\end{eqnarray}
	%\begingroup\makeatletter\def\f@size{9}\check@mathfonts
	%\def\maketag@@@#1{\hbox{\m@th\large\normalfont#1}}%
	%	\begin{align}
\notag\\
	&U_{1}=\left(
			\begin{array}{cc}
			u & u' \\
			u' & u 
			\end{array}
		\right), \quad
	U_{2}=\left(
			\begin{array}{cc}
			u & u'\omega^{-\xi} \\
			u'\omega^{+\xi} & u
			\end{array}
		\right), 
		\notag\\
	&U_{3}=\left(
			\begin{array}{cc}
			u & u'\omega^{+\xi} \\
			u'\omega^{-\xi} & u
			\end{array}
		\right). \nn \\
	\end{align} %\endgroup
	%where $\bm{q}_{j}$ is the vectors which are from layer2's K point to layer1's K points (See Fig.\ref{Fig1_BZ}),
	The parameters $u=79.7$ meV and $u'=95.7$ meV are interlayer coupling strength between AA/BB and AB/BA stack region, respectively. The difference between $u$ and $u'$ effectively arise from the in-plane lattice relaxation and from the out-of-plane corrugation effect	\cite{PhysRevX.8.031087, PhysRevB.101.195425}.
	%where the difference between $u$ and $u'$ takes account of the lattice relaxation effect. 
The interlayer matrix $U$ depends on the strain via $\bm{q}_j$'s [Eq.~\eqref{eq_q2}].
%    The detailed calculation is presented in Appendix~\ref{sec_app2}.

%	We ignore the effect of moir\'e distortion on these interlayer coupling parameters u and u' because even we include this effect, the band structure is almost same with the result without it.

%	$u_{ij}^{(l)} = (\partial_{i} u_{j}^{(l)} + \partial_{j}u_{i})/2$ is the strain tensor defined by the 
%	displacement vector $\bm{u}^{(l)}(\bm{r})$.

	%because distortion vector $\bm{u}^{(l)}(\bm{r}) = \left( R\left(\mp\theta\right)+\mathcal{E}^{(l)} \right)\bm{r}$.
	%Thus the strain-induced vector potential depends on only anistropic tensile and shear elements.
	%In the next section, we show that the difference of each layer's strain-induced vector potential $\Delta\bm{A}_{dis} =\bm{A}_{\text{dis}}^{(1)}-\bm{A}_{\text{dis}}^{(2)}= \left(3 \beta\gamma_{0}/4ev\right)\left(2\Delta\epsilon_{-}, -\Delta\epsilon_{xy}\right)$ split the flat band along the energy axis,
	%where $\Delta\epsilon_{-} = \epsilon_{-}^{(1)} - \epsilon_{-}^{(2)}$ and $\Delta\epsilon_{xy} = \epsilon_{xy}^{(1)} - \epsilon_{xy}^{(2)}$.
 
 %%%%%
  
%%%%%%%%%%%%%%%%%%%%%%%%%%%%%%%%%%%%%%%%%%%%%%%%%%%%%%%%%%%%%%%%%%%%%%%%%%%%%%%%%%%%%%%%%%%%%%%%%%%%%%%%%%%%%%%%%%%%%%%
%\subsection{Band calculation}

\begin{figure*}
  \begin{center}
    \leavevmode\includegraphics[width=1. \hsize]{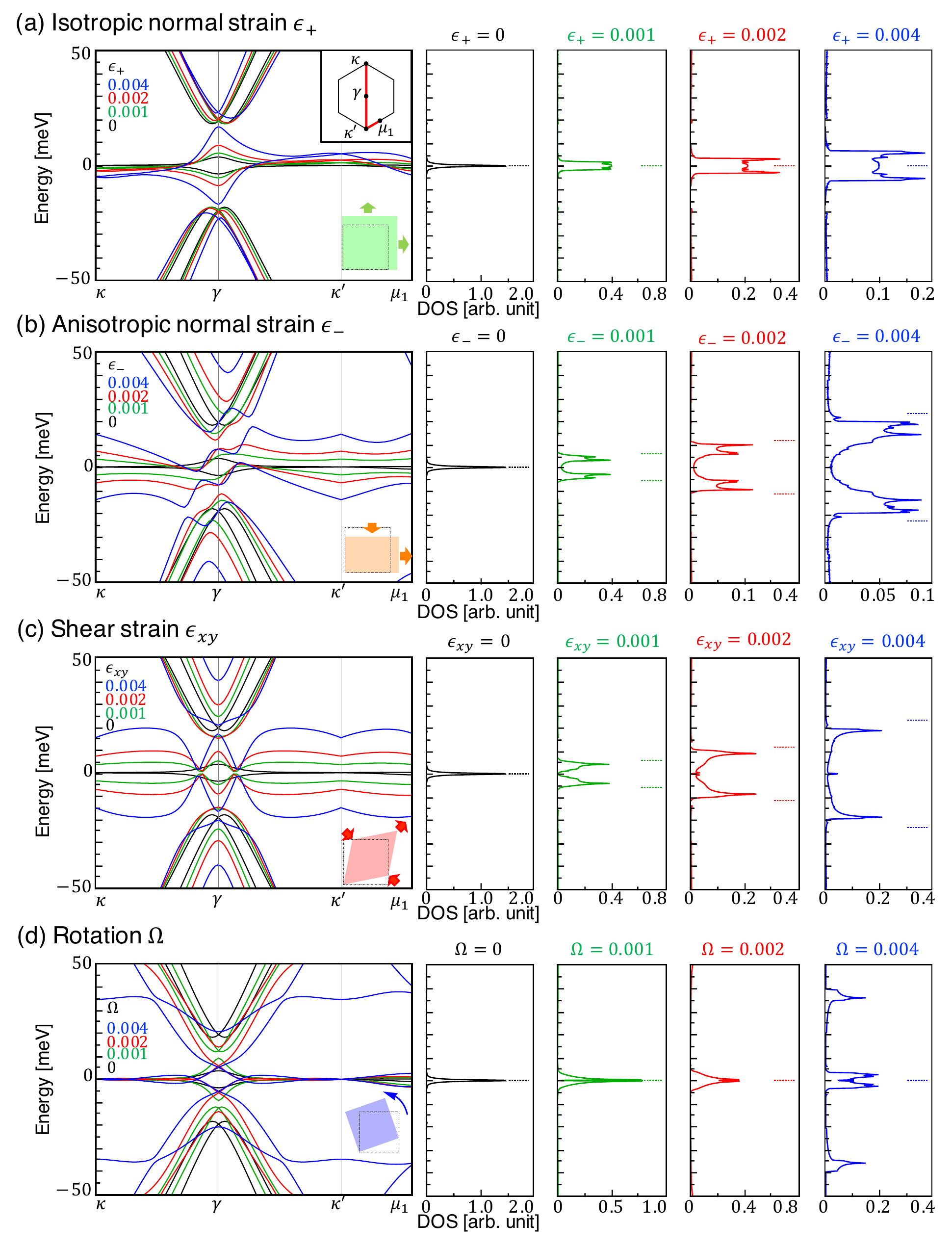}
    \caption{
   Band structure and the DOS of uniformly distorted magic-angle TBGs  with different types of strain components, $\epsilon_{+}, \epsilon_{-}, \epsilon_{xy}, \Omega$.
Different colors represent different amplitudes of strain.
 horizontal lines in the right panels (DOS) indicate energies of the split levels in the pseudo Landau level picture.
%  
 %   The band structure and DOS of the uniform model with each distortion elements.
  %  (a), (b), (c) and (d) represent the dependence of isotropic normal strain, anisotropic normal strain, shear strain and rotation respectively.
   % In the each row, the left panel is the band structure.
    %Black, Green, Red, Blue lines represent the result of the uniform model with each value of the interlayer difference of the distortion element are 0.0, 0.001, 0.002 and 0.004, respectively.
    %The perpendicular axis is the energy, and the horizontal is along the pass that is showed in the left figure of (a). 
    %The forth figures on the right are DOS(thick line) and the splitting of the PLL model(dot line), each color are corresponded the figure of the band structure.
    %The perpendicular is the energy and the horizontal axis is the DOS.
    %In the case of the isotropic normal strain(a) and rotation(d), the flat band still remain, while the anisotropic normal strain and shear strain split the flat band.
    %This split width of the flat band represent the good agreement with the result of the PLL model(dot line).
    }
    \label{Fig2_DOS_each_elements}
  \end{center}
  \end{figure*}

Below we investigate the effect of lattice distortion on the energy bands using the effective Hamiltonian, Eq.~\eqref{eq:TBGHamiltonian}.
In fact,  the electronic structure is  mainly affected by the interlayer asymmetric components of the strain tensor [Eq.~\eqref{eq_Delta_eps}]
,
%For later arguments, we introduce the interlayer difference of the strain components as
%    \begin{align}\label{eq_Delta_eps}
%    &\Delta \epsilon_{\pm} = \epsilon_{\pm}^{(1)}-\epsilon_{\pm}^{(2)},
%    \notag\\
%    &\Delta \epsilon_{xy} = \epsilon_{xy}^{(1)}-\epsilon_{xy}^{(2)},
%    \notag\\
%    &\Delta \Omega = \Omega^{(1)}-\Omega^{(2)},
%    \end{align}
and in particular, the flat band is highly sensitive to $ \epsilon_{-}$ and $ \epsilon_{xy}$.
To demonstrate this, we calculate the energy bands of 
the magic-angle TBG $(\theta=1.05^\circ)$ in the presence of asymmetric strain $\mathcal{E}^{(1)} = - \mathcal{E}^{(2)} =  \mathcal{E}/2$,
where different types of strain components $ \Omega, 
 \epsilon_{+},  \epsilon_{-},  \epsilon_{xy}$ are considered independently.
Figure \ref{Fig2_DOS_each_elements} shows the 
band dispersion and the density of state (DOS) 
in individual strain components, where black, green, red, and blue lines represent the strain amplitude (i.e., value of $ \Omega, 
 \epsilon_{+},  \epsilon_{-},  \epsilon_{xy}$) of $0, 0.001, 0.002$ and $0.004$, respectively.

We clearly observe that 
the central flat band is particularly sentsitive to $ \epsilon_{-}$ and $ \epsilon_{xy}$,
where a small distortion of 0.001 leads to a significant split of the flat band about 20 meV.
In constrast, $ \epsilon_{+}$ and $ \Omega$ gives relatively minor effects. 
%compared to $\Delta \epsilon_{-}$ and $\Delta \epsilon_{xy}$ of the same magnitude.
$ \epsilon_{+}$ moves the Dirac points at $\kappa$ and $\kappa'$ in the opposite directions in energy, resulting in a smaller DOS split.
$\Omega$ shifts the twist angle from the magic angle
and slightly broadens the flat band.
The strain-induced flat band splitting was also found the previous work, which considered the effect of uniaxial heterostrain in TBG \cite{PhysRevB.100.035448,xie2019spectroscopic,PhysRevB.104.045403,PhysRevResearch.4.013209},
which corresponds to $\epsilon_{-}$ and $\epsilon_{xy}$ in our notation.

It should also be noted that
the split flat bands in Fig.~\ref{Fig2_DOS_each_elements} are not completely separated, but stick together at certain points near $\gamma$ (off the path shown in Fig.~\ref{Fig2_DOS_each_elements}) \cite{PhysRevB.100.035448}.
These Dirac points are originally located at $\kappa$ and $\kappa'$ in the non-distorted TBG,
and when a uniform distortion is applied, they move without gap opening
under the protection of the $C_{2z}T$ symmetry. The two Dirac points cannot pair-annihilate because they have the same Berry phase \cite{PhysRevResearch.3.013033}.

\subsection{Pseudo Landau Level approximation}

As shown in the previous section, the flat band is split significantly by anisotropic normal strain $\epsilon_{-}$ and shear strain $\epsilon_{xy}$, while not much by other components.
We explain this by using the pseudo Landau level picture of TBG \cite{PhysRevB.99.155415},
which describes the flat band as the Landau level (LL) under a moir\'{e}-induced fictitious magnetic field.
We apply the same formulation to the strained TBG, Eq.~\eqref{eq:TBGHamiltonian},
and analytically estimate the flat-band split energy.

The pseudo-LL Hamiltonian is obtained by rewriting the Hamiltonian matrix [Eq.~\eqref{eq:TBGHamiltonian}]
in the basis $(\psi^{+}_{A}, \psi^{+}_{B}, \psi^{-}_{A}, \psi^{-}_{B})$ where $\psi^{\pm}_{X} = (\psi^{(1)}_{X} \pm \mathrm{i}\psi^{(2)}_{X})/\sqrt{2}$,
and then expanding it in $\bm{r}$ 
with respect to the origin (the AA-point)
upto the first order \cite{PhysRevB.99.155415}.
We ignore $\left(R\left(\mp\theta/2\right)+\mathcal{E}^{(l)} \right)^{-1}$ in Eq.~\eqref{eq:monolayerGrapheneHamiltonian},
which gives only higher order effects.
The detailed calculation is presented in Appendix~\ref{sec_app3}.
%	As we show in the above section, the flat band split by only anisotropic normal strain and shear strain.
%	To understand the origin of this double peak, we use the pseudo Landau level(PLL) model of TBG\cite{PhysRevB.99.155415}.
%	By using this model, we get the PLL hamiltonian with distortion as follows(Appendix A),

As a result, the effective Hamiltonian is written as
	\begin{align} \label{eq:PseudoModel}
	H_{\rm PLL} =
	    \left(
			\begin{array} {cc}
			H_{+} & V^{\dagger} \\
			V & H_{-}
			\end{array}
		\right),
	\end{align}
where
	\begin{align} \label{eq:PseudoModeldiagonal}
	&H_{\pm} = -\hbar v \left(\bm{k}\pm\frac{e}{\hbar}\bm{a}(\bm{r})\right)\cdot \bm{\sigma},
	\\
	\label{eq:PLL_vector_potential}
	&\bm{a}(\bm{r}) = \xi
	\frac{2\pi u'}{eva}
	(\theta-\Omega)
		\left(
			\begin{array} {c}
			-y \\
			x
			\end{array}
			\right).
\end{align} 
Eq.~\eqref{eq:PseudoModeldiagonal} is essentially the Dirac Hamiltonian
under a uniform magnetic field 
%$\pm\bm{B}= \pm \nabla \times \bm{a}(\bm{r}) =$ 
$\nabla \times (\pm \bm{a}) = (0,0,\pm b_{\rm eff})$ with $b_{\rm eff}=\xi[4\pi u'/(e^2va)](\theta-\Omega)$.
Note that the pseudo vector potential $\bm{a}(\bm{r})$ originates from the inter-sublattice coupling $u'$ in the 
moir\'{e} interlayer Hamiltonian [Eq.~\eqref{eq:Moire_interaction}],
and it should be distinguished from the strain-induced vector potential $\bm{A}^{(l)}$.

The off-diagonal matrix $V$ is given by
\begin{equation} \label{eq:PseudoModelpotential}
V = \left(-3\mathrm{i} u I_{2}- \frac{e v}{2}\bm{A}\cdot\bm{\sigma}\right)\e^{-\i\frac{2e}{\hbar}\chi(\bm{r})},
\end{equation}
where $I_{2}$ is a $2\times 2$ identity matrix,
$u$ is the intra-sublattice coupling in moir\'{e} interlayer Hamiltonian [Eq.~\eqref{eq:Moire_interaction}], and
\begin{align}		
	&\bm{A}=\bm{A}^{(1)}-\bm{A}^{(2)}=
	\xi \frac{3}{2}\frac{\beta\gamma_{0}}{ev}
	\begin{pmatrix}
	\epsilon_{-}\\
	-\epsilon_{xy}
	\end{pmatrix},
	\label{eq:Delta_A}
	\\
	&\chi(\bm{r}) = \xi
	\frac{\pi u'}{eva}
	\left[(x^2+y^2)\epsilon_{+}
	+(x^2-y^2)\epsilon_{-}
	+xy\epsilon_{xy}
	\right]. \label{eq:PLL_gauge_potential}
%		&\chi(\bm{r}) = \xi
%	\frac{2\pi u'}{eva}
%	\left(\Delta\epsilon_{+}\frac{x^{2}+y^{2}}{2}+\Delta\epsilon_{-}\frac{x^{2}-y^{2}}{2}+\Delta\epsilon_{xy}\frac{xy}{2}\right).
\end{align} 
Here $\bm{A}^{(l)}$ is the strain-induced vector potential
argued in the previous section. 

In the absence of the off-diagonal matrix $V$, the eigenstates are given by the pseudo LLs of sector $H_\pm$.
%in the pseudo magnetic field $\pm b_{\rm eff}$.
%The flat band of TBG corresponds to the 0th pseudo LLs.
For $\xi=+$ valley, it is explicitly written as
\begin{equation}\label{eq:pseudo_LL}
    |+,0, m\rangle = 
    \begin{pmatrix}
    0\\ \varphi_{0,m} \\0 \\ 0
    \end{pmatrix},
    \quad
    |-,0, m\rangle = 
    \begin{pmatrix}
    0 \\0 \\ \varphi_{0,m} \\ 0
    \end{pmatrix},
\end{equation}
where $\varphi_{0,m}(\bm{r}) \propto e^{-im\phi} e^{-r^2/(4l_{\rm eff}^2)}$ is the 0th LL wavefunction
with angular momentum $m$
expressed in the polar coordinate
$\bm{r} = r(\cos\phi, \sin\phi)$,
and $l_{\rm eff}= \sqrt{\hbar/(eb_{\rm eff})}$.
%Since the Dirac Hamiltonians $H_\pm$ have opposite pseudo magnetic fields, the 0th  LLs have exactly opposite sublattice polarization as seen in Eq.~\eqref{eq:pseudo_LL}.
The 0th  LLs in Eq.~\eqref{eq:pseudo_LL} have exactly opposite sublattice polarization (i.e., $|+,0, m\rangle$ on B, and $|-,0, m\rangle$ on A),
because the Dirac Hamiltonians $H_\pm$ have opposite pseudo magnetic fields
$\pm b_{\rm eff}$.

In the absence of distortion ($\bm{A} = \chi = 0$),
the 0th LLs remain the zero-energy eigenstates
even we include the off-diagonal terms $- 3 \mathrm{i} u I_{2}$ [Eq.~\eqref{eq:PseudoModelpotential}], 
because $I_{2}$ does not mix different sublattices.
The flat band of TBG is understood by these degenerate 0th LLs.
Since the effective Hamiltonian Eq.~\eqref{eq:pseudo_LL} is based on the linear expansion around $\bm{r}=0$ (the AA spot), the approximation is valid for the LL wavefunctions with small angular momenta $m$'s, which are well localized to $\bm{r}=0$.

When we switch on the disortion terms,
the 0th Landau levels are immediately hybridized by $\bm{A}\cdot\bm{\sigma}$
in the off-diagonal matrix $V$, and split into $E = \pm \Delta E/2$, where
\begin{equation}\label{eq:Delta_E}
\Delta E = e v |\bm{A}|
   = \frac{3}{2}\beta\gamma_{0}\sqrt{\epsilon_{-}^2 + \epsilon_{xy}^2}.
%\label{eq:Delta_E}
%    E_\pm \approx \pm \frac{e v}{2} |\Delta\bm{A}|
 %   =
  % \pm \frac{3}{4}\beta\gamma_{0}
%	\sqrt{\Delta\epsilon_{-}^2 + \Delta\epsilon_{xy}^2}.
\end{equation}
Note that the pseudo gauge potential $\chi(\bm{r})$ only contributes to the phase factor of the coupling matrix elements [Eq.~\eqref{eq:PseudoModelpotential}], giving a higher order correction to the splitting energy (see, Appendix \ref{sec_app3}).
Eq.~\eqref{eq:Delta_E} explains the exclusive dependence of the flat band splitting on $\epsilon_{-}$ and $\epsilon_{xy}$.
Considering $(3/2)\beta \gamma_0\approx$ 13\,eV,
%$(3/4)\beta \gamma_0\approx$ 6.4eV,
a distortion ($\epsilon_{-},  \epsilon_{xy})$ of the order of $10^{-3}$ corresponds to a split width $\Delta E \sim$ 10 meV.

In Fig.~\ref{Fig2_DOS_each_elements}, 
horizontal red lines represent $\pm \Delta E/2$ of Eq.~\eqref{eq:Delta_E},
showing a good agreement with the actual split width of the DOS.
In the energy bands, the structures at $\kappa$, $\kappa'$ and $\mu_i$ are nicely explained by this simple splitting picture. On the other hand, the energy bands around $\gamma$ point is rather complicated and cannot be captured by the same approximation. This is consistent with the fact that the wavefunction at $\gamma$ is extended over the entire moir\'{e} pattern unlike those at $\kappa$, $\kappa'$ and $\mu_i$ concentrating on AA points \cite{PhysRevB.98.235158,PhysRevResearch.1.033072,PhysRevB.102.155149,Nguyen_2021},
and hence the pseudo LL approximation (assuming the localization at AA point) fails. 
The Dirac band touching mentioned above actually occurs near $\gamma$.

\section{TBG with non-uniform distortion}
\label{sec:n-um}

%In this section, we discuss about the effect of the non-uniform distortion that are observed in the experiment recently.
%We show the local flat band at AA stack region split by the local strain-induced vector potential like uniform model, due to the localization of the flat band, structure of local flat band is understood by uniform approximation even wave length of distortion is few times moire lattice constant.
%Finally, we also show the double peak of DOS is broken by the large distortion.

%%%%%%%%%%%%%%%%%%%%%%%%%%%%%%%%%%%%%%%%%%%%%%%%%%%%%%%%%%%%%%%%%%%%%%%%%%%%%%%%%%%%%%%%%%%%%%%%%%%%%%%%%%%%%%%%%%%%%%%
\subsection{Theoretical modelling}

In this section, we construct a theoretical model to simulate
a non-uniform distortion in TBG.
We consider a super moir\'e unit cell composed of $n_{\rm SM}\times n_{\rm SM}$ 
original moir\'{e} units ($n_{\rm SM}$: integer), and assume that the lattice distortion is periodic with the super period as illustrated in Fig.~\ref{Fig3_Lattice_distortion}.
The primitive lattice vectors for the super unit cell
are given by $\bm{L}_{j}^{\rm SM} = n_{\rm SM}\bm{L}_{j}^{\rm M}$
and the corresponding reciprocal lattice vectors are 
$\bm{G}_{j}^{\rm SM} = \bm{G}_{j}^{\rm M}/n_{\rm SM}$.

We define the in-plane displacement vector of layer $l=1,2$ as
\begin{eqnarray} \label{def:distortionvector}
	\bm{u}^{(l)}(\bm{r}) = \sum_{\bm{p}} \bm{C}_{\bm{p}}^{(l)}\e^{-\left(\lambda\left|\bm{p}\right|/2\pi\right)^2} \e^{\i \bm{p}\cdot\bm{r}},
\end{eqnarray}
which represents a deformation such that a carbon atom of layer $l$
at a position $\bm{r}$ is shifted to $\bm{r} + \bm{u}^{(l)}(\bm{r})$.
Here $\bm{p}$ runs over $\bm{p}=m_{1}\bm{G}^{\rm SM}_{1} + m_{2}\bm{G}^{\rm SM}_{2}$, 
and $\lambda$ is the characteristic wave length of the spatial dependence of $\bm{u}^{(l)}(\bm{r})$.
The amplitude $\bm{C}_{\bm{p}}^{(l)} = (C_{\bm{p},x}^{(l)}, C_{\bm{p},y}^{(l)})$ is a two-dimensional random vector which satisfy $\bm{C}_{\bm{-p}}^{(l)}=\bm{C}_{\bm{p}}^{(l) *}$ for real-valued $\bm{u}^{(l)}(\bm{r})$.
We assume that different components of $\bm{C}_{\bm{p}}^{(l)}$ are totally uncorrelated such that
\begin{equation}
\langle C_{\bm{p},i}^{(l)} C_{\bm{p}',j}^{(l')*} \rangle = 
\delta_{l,l'}\delta_{\bm{p},-\bm{p}'}\delta_{i,j} C_0^2,
\end{equation}
where $\langle\rangle$ is the sampling average and 
$C_0$ is a length parameter to characterize the amplitude of the random displacement field.

The local strain tensors and the rotation angle can be expressed in terms of $\bm{u}^{(l)}(\bm{r})$ as
\begin{align}
    &\epsilon^{(l)}_{ij}(\bm{r}) = \frac{1}{2}\left(\partial_{i}u^{(l)}_{j} + \partial_{j}u^{(l)}_{i}\right) \\
    &\Omega^{(l)}(\bm{r}) = \frac{1}{2}\left(\partial_{x}u^{(l)}_{y} - \partial_{y}u^{(l)}_{x}\right).
\end{align}
As in the uniform case,
we define $\epsilon_{\pm}^{(l)}(\bm{r})$
by Eq.~\eqref{eq_epsilon_pm},
%$\epsilon_{\pm}^{(l)}(\bm{r}) =[\epsilon_{xx}^{(l)}(\bm{r})\pm\epsilon_{yy}^{(l)}(\bm{r})]/2$, 
and relative strain components $\epsilon_{\pm}(\bm{r}), \epsilon_{xy}(\bm{r}), \Omega(\bm{r})$ by Eq.~\eqref{eq_Delta_eps}.
We introduce the magnitude of distortion, $\epsilon$, as the root mean square of the interlayer difference of the strain tensor elements [Eq.~\eqref{eq_Delta_eps}], or,
\begin{equation}
    \epsilon \equiv \sqrt{\langle |\epsilon_\pm|^{2} \rangle} = \sqrt{\langle |\epsilon_{xy}|^{2}\rangle} = \sqrt{\langle |\Omega|^{2}\rangle} = \sqrt{\frac{\pi^3}{2} \frac{C_0^2 S_{\rm SM}}{\lambda^4}},
\end{equation}
where $S_{\rm SM} = |\bm{L}_1^{\rm SM}\times \bm{L}_2^{\rm SM}|$ is the area of the super moir\'e unit cell.

Figure \ref{Fig3_Lattice_distortion} show examples of distorted moir\'e patterns in the magic-angle TBG($\theta = 1.05^\circ$) with different values of $\epsilon=$ 0, 0.0006, 0.0012, 0.0018, where $n_{\rm SM} = 8$ (indicated by a big parallelogram) and $\lambda=7L_{M}$.
We adopted a continuous color code to express the stacking sequence \cite{PhysRevB.96.075311},
where the bright region represents local AA stack and the dark region represents AB/BA stack. The red dots are the AA spots of the non-distorted TBG for reference.
It should be noted that a small distortion in graphene lattice of the order of $\epsilon$ is magnified to the moir\'e disorder of $\epsilon/\theta \sim 60\epsilon$.

We calculate the energy spectrum by using an extended continuum model incorporating non-uniform lattice distortion \cite{PhysRevB.101.195425}.
The Hamiltonian
is given by Eq.~\eqref{eq:TBGHamiltonian},
where the diagonal blocks are replaced by  
\begin{flalign} \label{eq:monolayerGrapheneHamiltonian_with_nonuniform}
    H_{l}(\bm{k})=-\hbar v \left(\bm{k}+\frac{e}{\hbar}\bm{A}^{(l)}(\bm{r})\right)\cdot \bm{\sigma},
\end{flalign}
with the local strain-induced vector potential
\begin{eqnarray}\label{eq_A1_Non_uniform}
	\bm{A}^{(l)}(\bm{r}) &=& \xi
	\frac{3}{2}\frac{\beta\gamma_{0}}{ev}
	\begin{pmatrix}
    \epsilon_{-}^{(l)}(\bm{r})\\
	-\epsilon_{xy}^{(l)}(\bm{r})\
	\end{pmatrix},
	%\frac{3}{4}\frac{\beta\gamma_{0}}{ev}
	%\begin{pmatrix}
    %\epsilon_{xx}^{(l)}(\bm{r})-\epsilon_{yy}^{(l)}(\bm{r})\\
	%-2\epsilon_{xy}^{(l)}(\bm{r})\
	%\end{pmatrix}.
\end{eqnarray}
and the interlayer coupling $U$ is replaced with,
%\mage{
\begin{align} \label{eq:Moire_interaction_non_uni}
&U= \sum_{j=1}^{3} U_{j}\, 
\e^{
\i \xi [
\bm{q}_{j} \cdot \bm{r}
\,+\, \bm{K}_{j}\cdot
(\bm{u}^{(2)}(\bm{r}) - \bm{u}^{(1)}(\bm{r}))
]
}.
\end{align}
Here $U_{j}$ are defined in Eq.~\eqref{eq:Moire_interaction}, 
$\bm{K}_{j}$ are the corner points of an intrinsic graphene [Eq.~\eqref{eq_K_graphene}]
and $\bm{q}_j$ are interlayer corner-point shifts [Eq.~\eqref{eq_q}] of non-distorted TBG.
In the diagonal matrix, we neglected the rotation matrix $\left(R\left(\mp\theta\right)+\mathcal{E}^{(l)} \right)^{-1}$ in Eq.~\eqref{eq:monolayerGrapheneHamiltonian},
which gives a minor effect in the uniform distortion case.

While in this paper we focus on the in-plane components of lattice displacement, real TBG samples also contain out-of-plane corrugations \cite{uchida2014atomic,van2015relaxation,lin2018shear}.
The primary effect of the corrugation is to differentiate the lattice spacing of AA-stacking and AB-stacking regions, which is effectively incorporated by the difference between $u$ and $u'$ parameters in the matrix $U$ \cite{PhysRevX.8.031087,PhysRevB.101.195425},
as already mentioned. We may also have an additional effect from non-uniform corrugation, which is left for future work.

\subsection{Energy spectrum and flat-band splitting}

\begin{figure*}
  \begin{center}
    \leavevmode\includegraphics[width=1. \hsize]{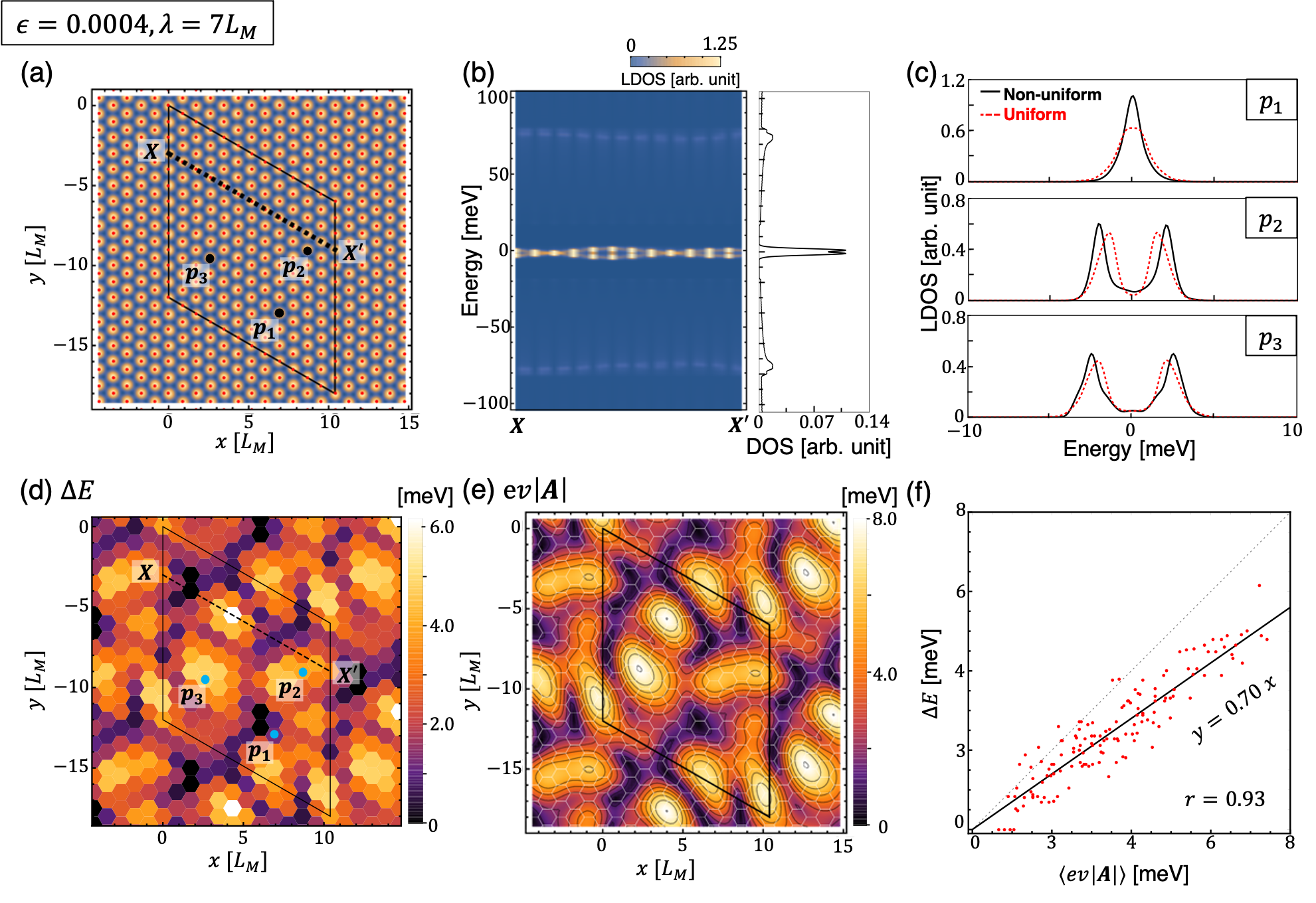}
    \caption{
    (a) Moir\'e pattern of a disordered magic-angle TBG with $\epsilon=0.0004, \lambda = 7L_{\rm M}$.
    The distortion is observed as slight shifts of AA points (yellow spots) relative to the regular red dots.
    (b) LDOS along line $XX'$ [defined by a broken line in (a)].
    (c) (Black, solid) LDOS at the points of $p_1, p_2, p_3$ in (a).
    (Red, dashed) LDOS at the AA point of the corresponding uniform TBG with the strain tensors fixed to the local value. 
    (d) The spatial distribution of the splitting energy $\Delta E$, or the energy distance between the two LDOS peaks. A hexagonal tile corresponds to a single moir\'e unit cell, and its color represents $\Delta E$ at the center of the hexagon (the AA point).
    (e) A contour plot of the interlayer difference of the strain-induced vector potential, $ev\left|\bm{A}(\bm{r})\right|$}. 
    (f) A scattered plot of $\Delta E$ and $ev\left|\bm{A}\right|$ (averaged in every moir\'e unit cell).
%    \red{
 %   (a) The moir\'e structure considered, where the distortion is barely observed as slight shifts of AA points (yellow spots) with respect to the regular red dots.
  %  (b) The LDOS along line $XX'$, which is defined by a broken line in Fig.~(a).
   % (c) The LDOS of the non-uniform TBG at the points of $p_1, p_2, p_3$ in Fig.~(a),and the density of states of the corresponding uniform TBGs. 
    %(d) The     spatial distribution of the splitting energy $\Delta E$, which is defined by the energy distance between the two LDOS peaks. Here a hexagonal tile corresponds to a single moir\'e unit cell, and its color represents $\Delta E$ at the center of the hexagon (the AA point).
    %(e) A contour plot of $ev\left|\Delta\bm{A}(\bm{r})\right|$.
    %(f) A scattered plot of $\Delta E$ and $ev\left|\Delta\bm{A}\right|$ (averaged in every moir\'e unit cell).
    %}
    %}
    \label{Fig4_Non_uniform_result}
  \end{center}
  \end{figure*}

Using the model obtained above, we calculate the local density of states (LDOS) for the magic-angle TBG ($\theta=1.05^\circ$) with a randomly-generated displacement configuration $\bm{u}^{(l)}(\bm{r})$.
First, we take $\epsilon = 0.0004$, $\lambda = 7L_{M}$, and $n_{\rm SM} = 12$.
Figure \ref{Fig4_Non_uniform_result}(a) illustrates the moir\'e structure, where the distortion is barely observed as a slight shift of AA points (yellow spots) with respect to the regular red dots.
In Fig.~\ref{Fig4_Non_uniform_result}(b), we plot the LDOS along line $XX'$, which is defined by a broken line in Fig.~\ref{Fig4_Non_uniform_result}(a).
%Here the horizontal axis is the position from $X$ to $X'$ and the vertical axis is the electron energy.
%The right figure plots the total density of states integrated over the super unit cell.
We can see that the LDOS of the flat band separates into upper and lower parts by a splitting energy depending on the position. This is quite different from the case of a random electrostatic potential which simply broadens the band width. 
%We define the local splitting energy $\Delta E(\bm{r})$ by
%the energy distance between the two LDOS peaks at the position $\bm{r}$. 
%The spatial distribution of $\Delta E$ in the super cell is shown in 
%Figure \ref{Fig4_Non_uniform_result}(d), where a hexagonal tile corresponds to a %single moir\'e unit cell, and its color represents $\Delta E$ at the AA point (the center of the hexagon) of the cell.
Figure \ref{Fig4_Non_uniform_result}(d) shows the spatial distribution of the splitting energy $\Delta E$, which is defined by the energy distance between the
two LDOS peaks. Here a hexagonal tile corresponds to a single moir\'e unit cell, and its color represents $\Delta E$ at the center of the hexagon (the AA point).

%In the recent experiments, few times moir\'e scale distortion is observed.
%So, we condisder the magic angle TBG($\theta=1.05$ degree) with the distorttion that has the period in the 12 times super moir\'e lattice.
%Fig.~\ref{Fig4_Non_uniform_result}(a) represents the moir\'e structure we calculated.
%The yellow and blue represent the AA and AB/BA stack region respectively.
%The black line is the super moir\'e unit cell, and the red points represent the AA stacking of the perfect moir\'e lattice.
%We took $\Delta \epsilon = 0.0004$ and $\lambda = 7L_{M}$.
%The moir\'e structure in Fig.~\ref{Fig4_Non_uniform_result}(a) looks almost normal moir\'e structure without distortion, however, the local flat band split by local strain elements as we show bellow.

%To discuss the structure of the local flat band, we calculate the local DOS (LDOS).
%The left figure of Fig.~\ref{Fig4_Non_uniform_result}(b) is the LDOS along the black dot line in of Fig.~\ref{Fig4_Non_uniform_result}(a) and the right figure represents the Total DOS.
%We take the Energy axis on the vertical and space position from X to X' on the horizontal.
%The yellow blight region show the higher LDOS value than others blue region.
%In the left figure of Fig.~\ref{Fig4_Non_uniform_result}(b), the split width of the local flat band at each AA stack depends on the position.

Actually, the local split width of the flat band is almost solely determined by
the local value of the interlayer difference of the strain-induced vector potential, 
\begin{equation}
\bm{A}(\bm{r})=\bm{A}^{(1)}(\bm{r})- \bm{A}^{(2)}(\bm{r}),
\end{equation}
and the local splitting energy is approximately given by $\Delta E \sim ev |\bm{A}(\bm{r})|$ as in the uniform case [Eq.~\eqref{eq:Delta_A}].
To demonstrate this, we show a contour plot of $ev\left|\bm{A}(\bm{r})\right|$ in Fig.~\ref{Fig4_Non_uniform_result}(e).
We observe a nearly perfect agreement with the distribution of $\Delta E$ in Fig.~\ref{Fig4_Non_uniform_result}(d).
We also present a scattered plot of $\Delta E$ and $ev\left|\bm{A}\right|$ (averaged in every moir\'e unit cell)
in Fig.~\ref{Fig4_Non_uniform_result}(f),
where we have a high correlation coefficient $r \approx 0.93$,
and a fitted line is given by $\Delta E \approx 0.7 ev|\bm{A}|$.
The strong correlation between the splitting width and  the strain-induced vector potential is a special property of the magic-angle flat band, as it relies on its peculiar Landau level like wavefunction.
On the other hand, the position of the satellite peaks (around $\pm 80$ meV in Fig.~\ref{Fig4_Non_uniform_result}) is totally uncorrelated with $ev\left|\bm{A}\right|$ (the correlation coefficient about $r\sim 0.1$),
but it is weakly correlated with the local twist angle $\Omega$  ($r\sim 0.5$). 

These results suggest that the local electronic structure in the flat band region of non-uniform TBG is well described by a uniform Hamiltonian with the strain tensors fixed to the local value.
In Fig.~\ref{Fig4_Non_uniform_result}(c), we plot the LDOS of the non-uniform TBG at the points of $p_1, p_2, p_3$ in Fig.~\ref{Fig4_Non_uniform_result}(a),
and the local density of states of the corresponding uniform TBGs at AA point. Indeed, we see a nice agreement between the two curves.
We also note that the LDOS is never completely gapped out at $E=0$, in accordance with the calculation of uniformly-strained TBGs where the two flat bands are always connected by the Dirac points.

\begin{figure*}
  \begin{center}
    \leavevmode\includegraphics[width=1. \hsize]{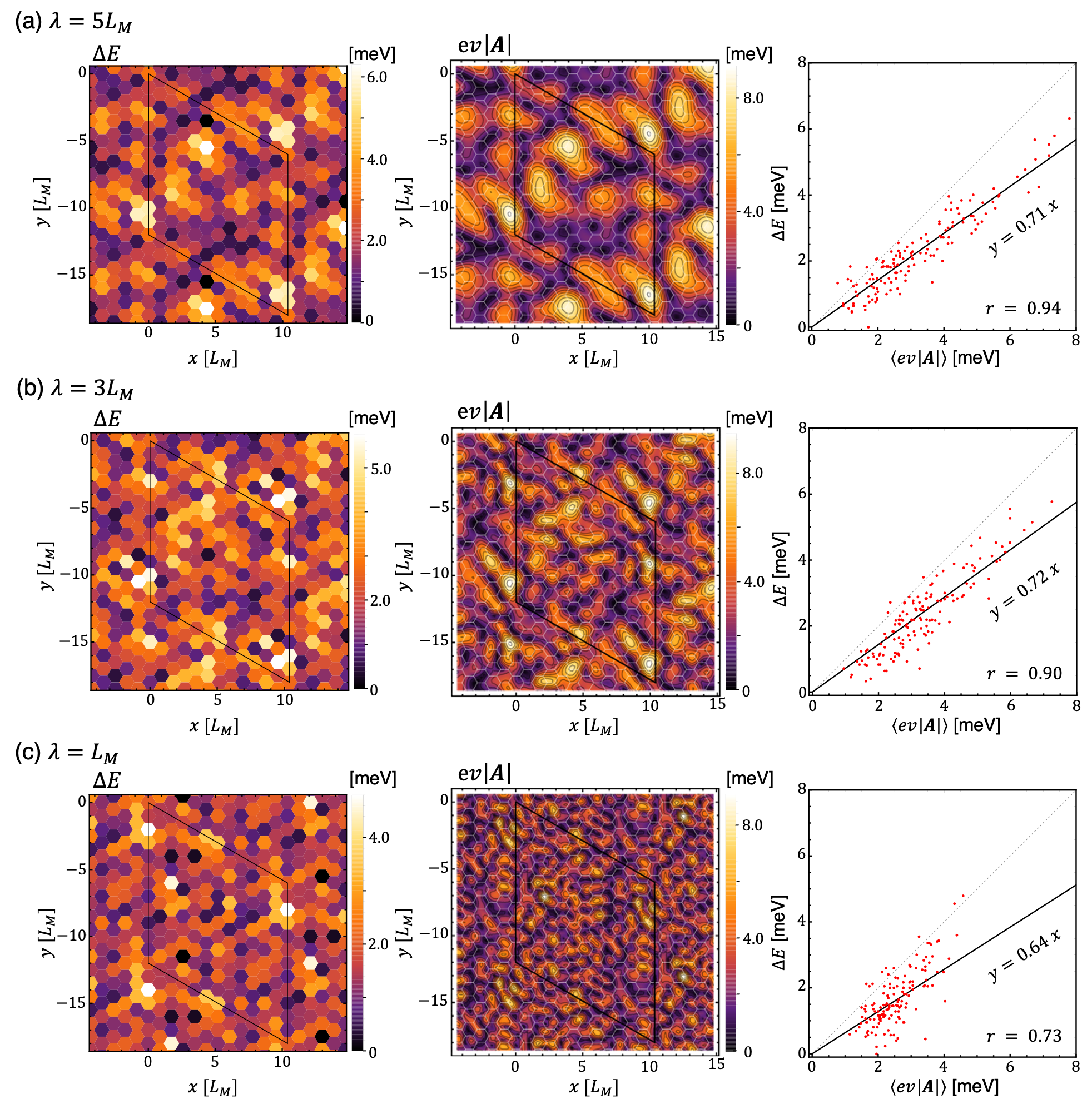}
    \caption{
    Plots similar to Figs.~\ref{Fig4_Non_uniform_result}(d)-(f) calculated for different characteristic wave lengths $\lambda =5L_M, 3L_M, L_M$.
    }
    %The wave length $\lambda$ dependence of the corresponding between the split width $\Delta E/2$ and the effectice vector potential $ev\left|\Delta\bm{A}(\bm{r})\right|/2$.
    %(a), (b) and (c) represent the case the wave length $\lambda = 5L_{M}$, $3L_{M}$ and $L_{M}$.
    %In the each figure, the left plot represents the split width $\Delta E/2$ at each AA stacking.
    %The middle is the strain-induced vector potential $ev\left|\Delta\bm{A}(\bm{r})\right|/2$ in the real space.
    %The horizontal axis and vertical axis are x and y axis respectively in the left and middle panels.
    %The black line is the super moir\'e unit cell.
    %The right is the scatter plot of the split width $\Delta E/2$ and the strain-induced vector potential avetaged in the each hexagon.
    %The black line is the regressionline and the gray dot line is the plot of Eq.~\eqref{eq:Delta_E}.
    %We write the correlation coefficient $r$ in the each figure.
    %The agreement between the split width $\Delta E/2$ and $ev\left|\Delta\bm{A}(\bm{r})\right|/2$ still remain the case the wave length $\lambda = L_{M}$.
    %}
    \label{Fig7_A_split_destribution}
  \end{center}
 \end{figure*}

The approximation with the local Hamiltonian is usually expected to be valid in a long-range limit with $\lambda \gg L_M$, but actually it works
fairly well down to a short-ranged distortion.
Figure \ref{Fig7_A_split_destribution} shows the plots similar to Fig.~\ref{Fig4_Non_uniform_result} calculated for different characteristic wave lengths, $\lambda =5L_M, 3L_M, L_M$.
The correlation coefficient between $\Delta E$ and $ev\left|\bm{A}\right|$ is found to be 0.90 at $\lambda = 3L_M$, and it is still 0.73 at $\lambda = L_M$.
We presume that it reflects the strongly localized feature of the flat-band wavefunctions.

	%% Gap out しないこと
	%As we discussed above, the band structure of TBG with uniform distortion still has Dirac points at charge neutral point.
	%So even in non-uniform case, local band structure does not gap out by distortion because of agreement between uniform and non-uniform models.
	%In fact, LDOS at AA stack region has finite value at CNP(Fig.~\ref{Fig4_Non_uniform_result}(c)).
	%this result show that distortion produces a double peak at AA stack region locally, while moir\'e distortion doesn't open the gap at CNP and TBG does not become an insulator due to moir\'e distortion.

%Actually, these split width decided by the interlayer difference of the local strain-induced vector potential $\Delta\bm{A}(\bm{r})$ like uniform distortion model.
%To understand this, we compare the distribution of $\Delta\bm{A}(\bm{r})$ and the local split width $\Delta E$.
%Fig.~\ref{Fig4_Non_uniform_result}(d) and (e) represent the distribution of $\Delta E$ and $ev\left|\Delta\bm{A}(\bm{r})\right|$, 
%and as we showed in Eq.~\eqref{eq:Delta_E}, these values are comparable directly.
%The black line is the super moir\'e unit cell in each figures.
%The blight region represnets the large $ev\left|\Delta\bm{A}(\bm{r})\right|$ and split width region, and these figures are good agreement.
%Thus, the split of the local flat band in the case of the non-uniform distortion. is also understand by the interlayer difference of the strain-induced vecotor potential.
%While, although the second peak of the energy band around $\pm 80$ meV(See Fig.~\ref{Fig4_Non_uniform_result}(b)) has position dependence due to distortion, it is uncorrelated with the strain-induced vector potential $ev\left|\Delta\bm{A}(\bm{r})\right|$.
  
  \begin{figure}
  %\begin{left}
    \leavevmode\includegraphics[width=1. \hsize]{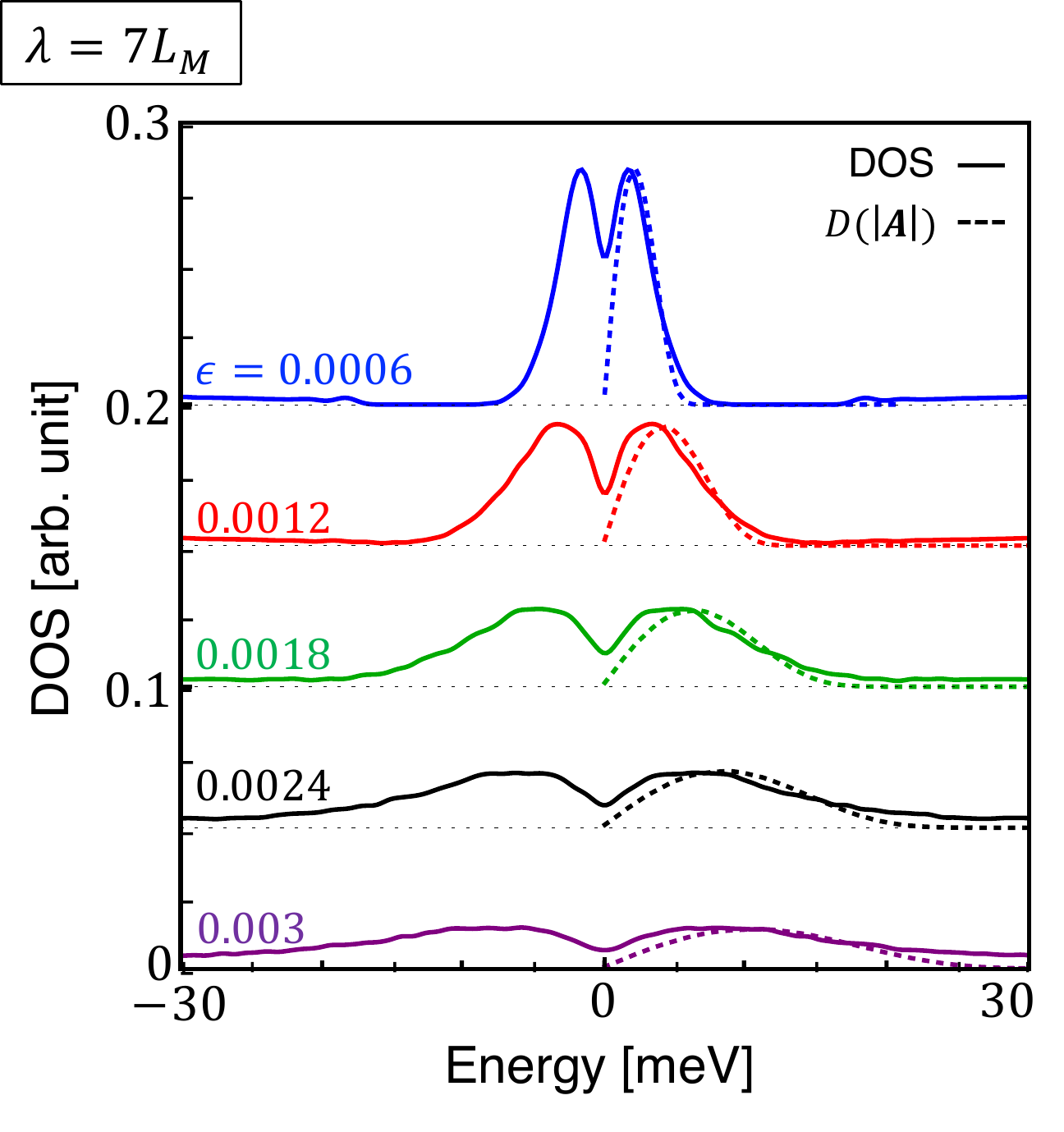}
    \caption{
    The total DOS of disordered magic-angle TBGs with different distortion amplitudes $\epsilon$.
    For each curve, we take an average over different random configurations.
    Broken lines are the distribution function $D(|\bm{A}|)$ with horizontal axis scaled by $E=0.7 ev|\bm{A}|$.
    }
    %\red{
    %Broken lines are the distribution function $D(|\Delta \bm{A}|)$ of the current strain model with each amplitude $\Delta \epsilon$.
    %We multiplied the horizontal axis of the $D(|\Delta \bm{A}|)$ plot by 0.7 to adjust to the DOS curve.}
    %\red{
    %(a) The total DOS of TBG with non-uniform distortion with different distortion amplitudes $\Delta \epsilon$.
    %For each curve, we take an overage over different random configurations.
    %(b) The distribution function  $D(|\Delta \bm{A}|)$ of the current strain model.
    %}
    %
    %
    %(a) This is the plot of the distribution function of $\left|\Delta\bm{A}\right|$ of Fig.~\ref{Fig4_Non_uniform_result}(e).
    %The vertical axis is the distribution function $D(ev\left|\Delta\bm{A}\right|)$, and horizontal is $ev\left|\Delta\bm{A}\right|$.
    %(b) The dependence of distortion strength$\Delta \epsilon$ on the total DOS of TBG at $\theta=1.05$ degree with non-uniform distortion.
    %We took $\lambda=7L_{M}$.
    %The vertical axis is DOS and horizontal is energy.
    %The color of line correspond the value in the figure.
    %The double peak is broken by distortion when the strength is more than $\Delta \epsilon = 0.0024$.
    %Also, as the strength of the distortion increases, it becomes broader for each of the two splitting peaks.
    \label{Fig5_distribution_A_and_TotalDOS}
  %\end{left}
  \end{figure}

Figure \ref{Fig5_distribution_A_and_TotalDOS} plots the total DOS of non-uniform TBG in different distortion amplitudes $\epsilon$ with $\lambda=7L_{M}$,
For each curve, we take an overage over different random configurations.
We see that the two-level splitting feature in the LDOS still remains as a double peak structure in the total DOS.
In increasing $\epsilon$, the curve is simply extended horizontally, as expected the relationship $\Delta E \sim ev|\bm{A}|$.
The form of the DOS curve is roughly determined by the distribution function  $D(|\bm{A}|)$,
which is plotted as broken line in Fig.~\ref{Fig5_distribution_A_and_TotalDOS}
for the current model.
Here we scale the horizontal axis by $E=0.7 ev|\bm{A}|$ in accordance with Fig.~\ref{Fig4_Non_uniform_result}(f).

By using the formula Eq.~\eqref{eq:Delta_E}, we can roughly estimate the flat band split energy in real TBG samples.
In a recent local measurement of the magic-angle TBG \cite{Nature.581.7806} has shown that the local twist angle varies from $\theta =1.05^\circ$ to $1.18^\circ$, which amounts to $\Omega \simeq 0.001$ (rad).
By assuming that the strain tensor elements, $\epsilon_{\pm}$, $\epsilon_{xy}$, $\Omega$ have comparable magnitudes, 
the typical value of the flat band split width on this sample is estimated at $\Delta E \simeq 10$ meV using Eq.~\eqref{eq:Delta_E}.
These results suggest that, in realistic magic-angle TBGs with non-uniform moir\'e disorder, the flat band is not actually a single band cluster but it splits by a sizable energy in most places.
It is consistent with the STM measurements of TBGs near the magic angle \cite{Kerelsky2019,Choi2019}, where a significant separation of the LDOS was observed. 
The local flat-band separation may also be responsible for the pronounced Landau fan at the charge neutral point which is commonly observed in the transport experiments \cite{cao2018_43,doi:10.1126/science.aav1910,lu2019superconductors,Nature.581.7806},
since the two separate bands are always touching as argued in Sec.~\ref{sec_um}.
The splitting of the flat band would affect the ground state properties in the presence of the electron-electron interaction, since the Hilbert space of the half-split flat band is different from the original full flat band.
	
While we focus on the strain effect in this calculation, the distortion of the moir\'e pattern should also give rise to a non-uniform electrostatic potential via an inhomogeneous charge distribution\cite{PhysRevB.98.235158,doi:10.1073/pnas.1810947115,PhysRevB.99.140302,PhysRevB.102.201107}.
We expect that the effect is roughly captured by including a local shift of the energy in the present calculation. At the filling factor $\nu=2$
(i.e. half-filling of the upper flat band),
for instance, the upper LDOS peak would be aligned to the Fermi energy without changing the local splitting width, to achieve the homogeneous electron density of $\nu=2$.
We leave a detailed calculation including the electrostatic potential for future works.

	%Finally, we estimate the effect of moir\'e distortion in the actual TBG.
	%According to the Uri's experiment\cite{Nature.581.7806}, local twist angle has the range from 1.05 to 1.18 degree.
	%It show the angle variation of this sample is around $\Delta \Omega \simeq 0.001$.
	%By assuming $\Delta \epsilon_{\pm}$, $\Delta \epsilon_{xy}$, $\Delta \Omega$ are comparable, 
	%we can estimate the maximum value of the split width on this sample is $\simeq 10$ meV by Eq.~\eqref{eq:Delta_E}.
	%So, The flat band is not a sigle band cluster but mostly double-split place by place in the actual TBG.
	
	%Finally, we estimate the effect of moir\'e distortion in the actual TBG.
	%According to the Uri's experiment\cite{Uri}, local twist angle has the range from 1.05 to 1.18 degree.
	%we assume that the middle angle 1.115 degree is twist angle without distortion, we estimate the max value of the twist angle distortion is 0.065 degree.
	%We aren't sure about others distortion elements, however, we can estimate the max value of the $ev\left|\Delta\bm{A}(\bm{r})\right|/2$ that is comparable to the split width of the local flat band[See Eq.\ref{eq:Delta_E}] is $10$ meV by assuming all distortion elements are same order.
	%So we guess that the distortion produces the double peak at local AA stack region in the actual TBG of Uri's research\cite{Uri} and the maximum value of these split width is about 10 meV.

\section{Conclusion}
\label{sec_con}
We have studied the electronic structure of the magic-angle TBG with non-uniform moir\'e distortion by using an extended continuum model.
We found that the local density of states of the flat band is split by the local interlayer difference of anisotropic normal strain $\epsilon_-$ and shear strain $\epsilon_{xy}$, while  isotropic strain $\epsilon_+$ and rotation $\Omega$ give relatively minor effects.
The splitting of the flat band can well be described by
a pseudo landau level picture for the magic-angle flat band,
and an analytical expression of the splitting energy is obtained [Eq.~(\ref{eq:Delta_E})].
The coincidence between the splitting energy of the LDOS and the local strain is maintained even in a short-ranged distortion with $\lambda \sim L_{M}$, reflecting a highly-localized feature of the flat band wave function.

%
%We studied the electronic structure of magic-angle TBG with non-uniform moiré distortion by using extended continuum model.
%We found that the interlayer difference of the strain-induced vector potential $\Delta \bm{A}$, caused by $\Delta \epsilon_-$ and $\Delta \epsilon_{xy}$, split the flat band.
%While, contrary to general expectations, $\Delta \epsilon_+$ and $\Delta \Omega$ give relatively minor effects.
%The coincidence between the splitting energy of the flat band and the local values of $\Delta \bm{A}(\bm{r})$ is maintained even in non-uniform distortion with $\lambda = L_{M}$ due to a highly-localized feature of the flat band wave function.
%By using a pseudo landau level picture for the magic-angle flat band \cite{PhysRevB.99.155415}, we analytically showed that the splitting energy is mainly determined by the Eq.~(\ref{eq:Delta_E})

\section*{Acknowledgments}

This work was supported in part by 
JSPS KAKENHI Grant Number JP20H01840, JP20H00127, JP21H05236, JP21H05232 and by JST CREST Grant Number JPMJCR20T3, Japan.

%%%%%%%%%%%%%%%%%%%%%%%%%%%%%%%%%%%%%%%%%%%%%%%%%%%%%%%%%%%%%%%%%%%%%%%%%%%%%%%%%%%%%%%%%%%%%%%%%%%%%%%%%%%%%%%%%%%%%%
\appendix

\section{Pseudo Landau Level Hamiltonian} \label{sec_app3}

In this appendix, we derive the pseudo landau level Hamiltonian Eq.~\eqref{eq:PseudoModel} by applying the method of Ref.~\cite{PhysRevB.99.155415} to the disordered TBG.
By defining 
\begin{equation}
  \psi^{\pm}_{X} = (\psi^{(1)}_{X} \pm \mathrm{i}\psi^{(2)}_{X})/\sqrt{2}  
  \quad (X=A,B),
\end{equation}
%We take new basis of the hamiltonian as follow
%\begin{align}
%    \begin{pmatrix}
%	\psi_{X}^{(+)}\\
%	\psi_{X}^{(-)}\
%	\end{pmatrix}
%	= \frac{1}{\sqrt{2}}
%	\left(
%		\begin{array} {cc}
%		1 & \i \\
%		1 & -\i
%	    \end{array}
%	\right)
%	\begin{pmatrix}
%	\psi_{X}^{(1)}\\
%	\psi_{X}^{(2)}\
%	\end{pmatrix},
%\end{align}
the Hamiltonian matrix of Eq.~\eqref{eq:TBGHamiltonian} is written
in the basis $(\psi^{+}_{A}, \psi^{+}_{B}, \psi^{-}_{A}, \psi^{-}_{B})$ 
as
\begin{align} \label{eq:PseudoModel_Appendix}
	H =
	\left(
		\begin{array} {cc}
		h_{+}+\displaystyle\frac{\i}{2}(U-U^{\dagger}) & h_{-}+\displaystyle\frac{\i}{2}(U+U^{\dagger}) \\
		h_{-}-\displaystyle\frac{\i}{2}(U+U^{\dagger}) & h_{+}-\displaystyle\frac{\i}{2}(U-U^{\dagger})
		\end{array}
	\right),
\end{align}
where 
\begin{align}
    h_{+} &= -\Biggl(\hbar v \bm{k} + ev \frac{\bm{A}^{(1)}+\bm{A}^{(2)}}{2} \Biggr)
    \cdot\bm{\sigma} \nn \\
    h_{-} &= - ev\frac{\bm{A}^{(1)}-\bm{A}^{(2)}}{2}\cdot\bm{\sigma}.
\end{align}
In the following, we neglect the homostrain component $\bm{A}^{(1)}+\bm{A}^{(2)}$, and 
focus on the heterostrain part $\bm{A} = \bm{A}^{(1)}-\bm{A}^{(2)}$.

%Because each diagonal term have constant term $\frac{\bm{A}^{(1)}+\bm{A}^{(2)}}{2}$, we can remove it by the gauge transformation. %$(\psi^{+}_{A}, \psi^{+}_{B}, \psi^{-}_{A}, \psi^{-}_{B}) \to \e^{-\i \frac{\bm{A}^{(1)}+\bm{A}^{(2)}}{2}\cdot\bm{r}} (\psi^{+}_{A}, \psi^{+}_{B}, \psi^{-}_{A}, \psi^{-}_{B})$.
%Thus, we can replace $h_{+} \to -\hbar v\bm{k}\cdot\bm{\sigma}$.

%In following, we use an approximation that focus only the flat band of TBG.
Since the wavefuncton of the flat band is localized around the AA region, we expand the interlayer coupling matrix $U(\bm{r})$ around the AA stacking point $(\bm{r}=0)$ to the linear order of $r/L_{M}$.
As a result, we have
\begin{align}
    \frac{U+U^{\dagger}}{2} &= \sum_{j=1}^{3}U_{j}\cos{\bm{q}_{j}\cdot\bm{r}} \approx 3uI_{2}
    \label{eq_U+Udag}
\\
    \i\frac{U-U^{\dagger}}{2} &= 
    \sum_{j=1}^{3}U_{j}\sin{\bm{q}_{j}\cdot\bm{r}}  \approx 
    \sum_{j=1}^{3}U_{j}\bm{q}_{j}\cdot\bm{r}.
    \label{eq_U-Udag}
    %\nn \\
    %&\approx \sum_{j=1}^{3}U_{j}\left(\i\bm{q}_{j}\cdot\bm{r}\right) \nn \\
    %\left(
	%	\begin{array} {cc}
	%	0 & u'\left(\bm{q}_{1}+\omega^{-\xi}\bm{q}_{2}+\omega^{+\xi}\bm{q}_{3}\right)\cdot\bm{r} \\
	%	u'\left(\bm{q}_{1}+\omega^{-\xi}\bm{q}_{2}+\omega^{+\xi}\bm{q}_{3}\right)\cdot\bm{r} & 0
	%	\end{array} 
	%\right) \nn \\
	%
%	&\approx - \hbar v \frac{e}{\hbar}\left(\bm{a}(\bm{r})+\nabla\chi(\bm{r})\right)\cdot\bm{\sigma},
\end{align}
%where we used a formula of distorted interlayer shift of the %corner points of BZ.,
%\begin{align}
%    \bm{q}_{j} &= 
 %   \left(\frac{4\pi}{3a}\right)\left[
 %   R\left(\phi_{j}\right)
 %   \begin{pmatrix}
 %   \Delta\epsilon_{+} \\ \theta-\Delta\Omega
 %   \end{pmatrix}
 %   + R\left(-\phi_{j}\right)
 %   \begin{pmatrix}
 %   \Delta\epsilon_{-} \\ \Delta\epsilon_{xy}
 %   \end{pmatrix}
 %   \right],\\
 %   \phi_j &= \frac{2\pi}{3}(j-1). \nn
%\end{align}
By using Eqs. \eqref{eq_U-Udag} and ~\eqref{eq_q2},
the diagonal part of the Hamiltonian (\ref{eq:PseudoModel_Appendix}) is written as
\begin{equation}
    h_{+}\pm\displaystyle\frac{\i}{2}(U-U^{\dagger})
    = -\hbar v\left[\bm{k}\pm\frac{e}{\hbar}\left(\bm{a}(\bm{r})+\nabla\chi(\bm{r})\right)\right]\cdot\bm{\sigma}
\end{equation}
where $\bm{a}(\bm{r})$ is the pseudo vector potential of Eq.~\eqref{eq:PLL_vector_potential} 
and the $\chi(\bm{r})$ is the gauge potential of Eq.~\eqref{eq:PLL_gauge_potential}.
%where $\bm{a}(\bm{r})$ is the pseudo vector potential[Eq.~\eqref{eq:PLL_vector_potential}] originated from the inter-sublattice coupling $u'$
%and the $\chi(\bm{r})$ is the gauge potential, Eq.~\eqref{eq:PLL_gauge_potential}.
Finally, the effective Hamiltonian Eq.~\eqref{eq:PseudoModel}
is obtained by applying a gauge transformation,
\begin{align}
    \begin{pmatrix}
	\tilde{\psi}_{X}^{(+)}\\
	\tilde{\psi}_{X}^{(-)}\
	\end{pmatrix}
	= 
	\left(
		\begin{array} {cc}
		\e^{-\i\frac{e}{\hbar}\chi} & 0 \\
		0 & \e^{+\i\frac{e}{\hbar}\chi}
	    \end{array}
	\right)
	\begin{pmatrix}
	\psi_{X}^{(+)}\\
	\psi_{X}^{(-)}\
	\end{pmatrix}.
\end{align}

The coupling matrix elements in the 0th LLs are given by
\begin{align}
    &\langle -,0, m'|V|+,0, m\rangle 
    = \frac{ev}{2}\bm{A}\cdot \bm{\sigma} 
    \langle \varphi_{0, m'}|\e^{-\i\frac{2e}{\hbar}\chi(\bm{r})}|\varphi_{0, m}\rangle \nn \\
    &\qquad \approx \frac{ev}{2}\bm{A}\cdot \bm{\sigma}  \left[\delta_{m,m'} -2\i\frac{e}{\hbar} \langle \varphi_{0, m'}|\chi(\bm{r})|\varphi_{0, m}\rangle\right].
\end{align}
Therefore, the gauge potential $\chi$ only contributes to a higher order correction in the 0th LL splitting.

\bibliography{reference_distortionTBG}
\end{document}